\documentclass[12pt,preprint]{aastex}
\usepackage{apjfonts}
\usepackage{wrapfig}
\usepackage{amsmath}
\usepackage{multicol}

\newcommand{\fnm}{\footnotemark}
\newcommand{\fnt}{\footnotetext}

\newcommand{\ergs}{erg s$^{-1}$}

\newcommand{\xmmnewton}{{\it XMM-Newton}}
\newcommand{\chandra}{{\it Chandra}}
\newcommand{\suzaku}{{\it Suzaku}}
\newcommand{\hst}{{\it HST}}

\newcommand{\asca}{{\it ASCA}}

\newcommand{\rosat}{{\it ROSAT}}

\newcommand{\bootes}{Bo\"{o}tes} 
 
\newcommand{\spitzer}{{\it Spitzer}}

\newcommand{\logn}{$\log{N}$-$\log{S}$}

\newcommand{\newsec}{\vspace{1.5ex}}

\begin{document}


\thispagestyle{empty}
{

\noindent{\Large \rm \bf BLACK HOLES THROUGH COSMIC TIME:

Exploring the distant X-ray Universe with

extragalactic \chandra\ surveys}

{\large \em Ryan C. Hickox} 

{\large \em Chandra Newsletter cover article, Winter 2009}} \\

\begin{figure}[!h]
\includegraphics[width=0.75\textwidth,angle=90]{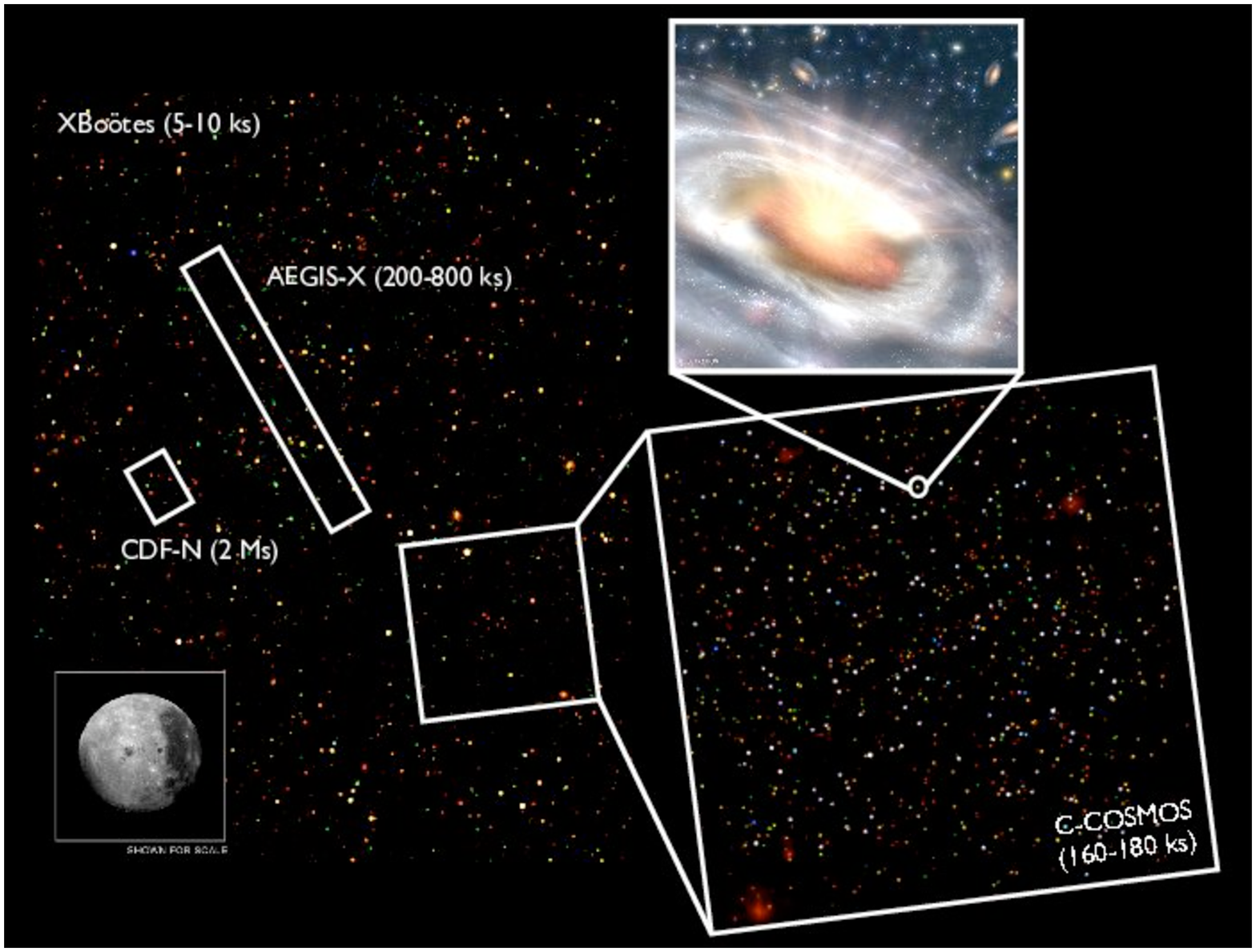}
\caption{\footnotesize X-ray images and survey areas for a few
representative \chandra\ surveys: X\bootes\ \citep{murr05}, C-COSMOS
\citep{elvi09sub}, AEGIS-X \citep{nand05egs}, and the Chandra Deep
Field-North \citep{alex03}.  The relative areas of each field are
superposed on the X\bootes\ image, and survey exposure times are
shown. (The three fields cover separate areas, and are shown together
only for comparison.)  The full C-COSMOS image is shown on an expanded
scale.  The inset illustration shows an active galactic nucleus; the
vast majority of the X-ray sources detected in \chandra\ surveys are
AGN.  {\em (Figure prepared by author, AGN illustration by
NASA/JPL-Caltech/T.Pyle-SSC.)}\label{fsurv}}
\end{figure}

In the last decade we have seen profound advances in our understanding
of the composition and evolution of the Universe.  Prominent among
these is the discovery that essentially all galaxies with stellar
bulges contain supermassive black holes (SMBH), which are believed to
be the relics of accretion in active galactic nuclei (AGN).  Further,
the masses of SMBHs are tightly correlated to the properties of their
host bulges, and  the energy released by AGN may have a
significant effect on the star formation history of galaxies. Thus it is
increasingly clear that growth and evolution of black holes and
galaxies are linked through cosmic time.

X-ray surveys are exceptionally powerful tools for studying the
evolution of black holes and their host galaxies, by detecting large
numbers of AGN over a wide range of redshifts and cosmic environments
from voids to groups and clusters.  With its superb angular
resolution, low background, and sensitivity in the energy range
0.5--8 keV, \chandra\ has been at the forefront of recent
extragalactic surveys.  In this article we provide an overview of some
of the leading \chandra\ surveys, and describe some recent results on
the composition of the cosmic X-ray background (CXB), the evolution of
black hole accretion, the nature of AGN populations, and links between
AGN and their host galaxies and environments.  This is an extremely
active and exciting field, with many key contributions made by
\xmmnewton, \suzaku, {\em INTEGRAL}, {\em Swift}, and other space and
ground-based observatories; here we will focus on just a few
representative results from \chandra\ surveys.

\newsec
\noindent {\bf Breadth and depth in \chandra\ surveys}

\begin{figure}[!t]
\begin{center}
\epsscale{0.7}
\plotone{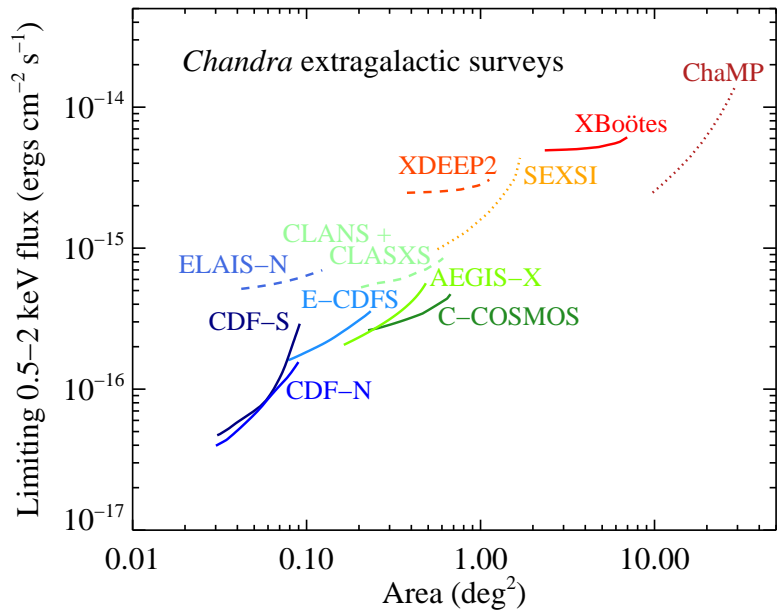}
\end{center}
\vspace{-6mm}
\caption{\footnotesize \chandra\ surveys span a wide range of depths
 and areas, in order to probe the widest possible ranges in redshift,
 luminosity, and environment.  The figure shows limiting 0.5--2 keV
 flux versus area for various \chandra\ blank-field and serendipitous
 extragalactic surveys (see text for references).  Since sensitivity
 varies across \chandra\ fields, for a given survey the area increases
 with increasing flux limit.  Lines show the sensitivity curves
 between 25\% and 75\% of the total area of each survey. Solid lines
 show contiguous surveys, dotted lines show serendipitous surveys, and
 dashed lines show surveys comprised of two or three separate fields
 with similar depths and multiwavelength coverage.  Flux limits are
 defined somewhat differently for different surveys, but generally
 correspond to a $\sim$50\% completeness limit.  For SEXSI (which is a
 hard X-ray selected serendipitous survey) the 2--10 keV limiting
 fluxes were converted from 0.5--2 keV by dividing by 6.5,
 corresponding roughly to the relative on-axis flux limits of the
 CDF-N in the soft and hard bands.  The AEGIS-X and X\bootes\
 sensitivities correspond to the 200 ks and 5 ks surveys,
respectively, while the sensitivity for the $\approx$10 ks XDEEP2
exposures is estimated from X\bootes. }
\label{farea}
\end{figure}

\begin{figure}[!t]
\begin{center}
\epsscale{0.55}
\plotone{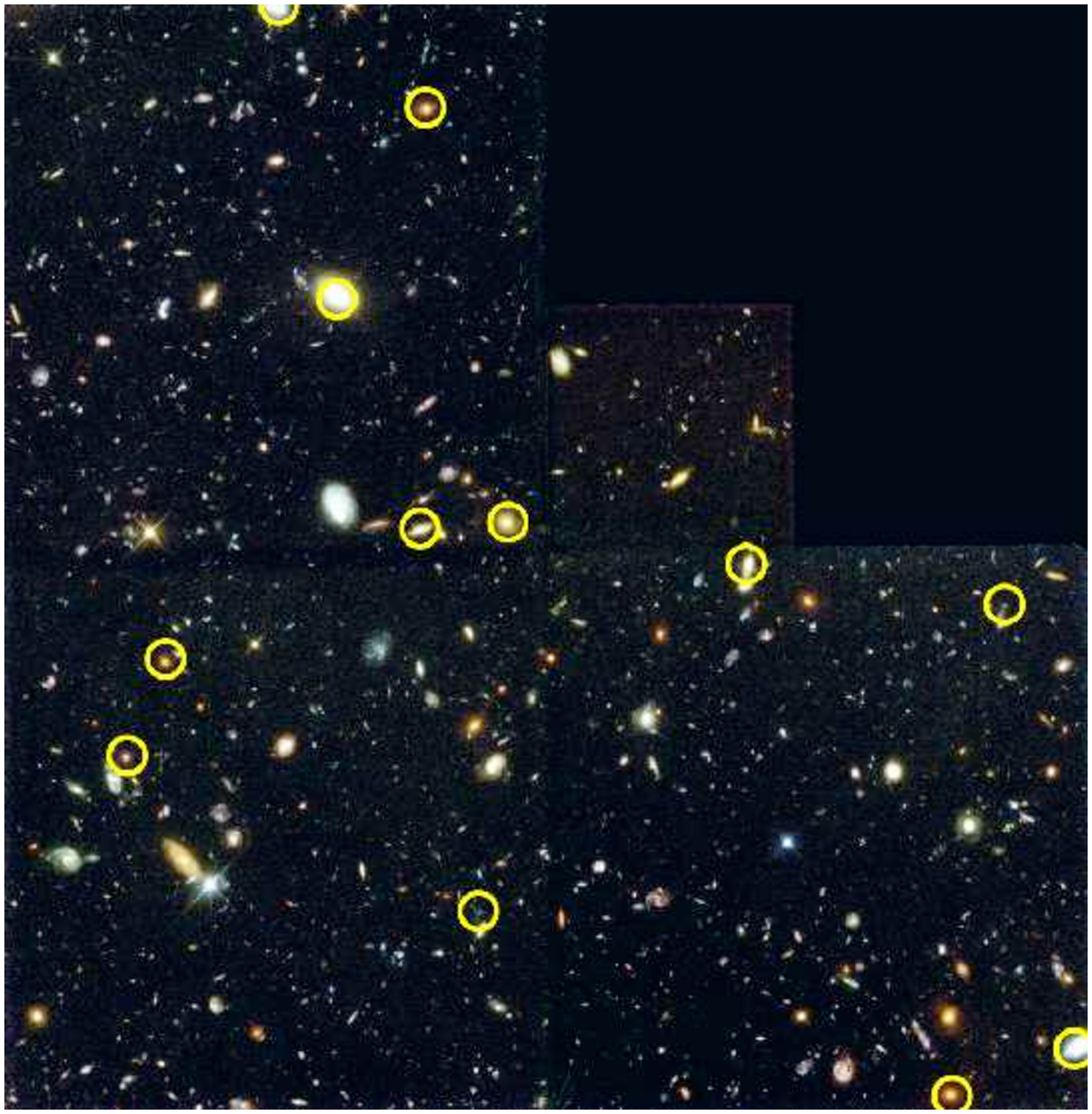}
\end{center}
\vspace{-6mm}
\caption{\footnotesize Optical counterparts of X-ray sources in the CDF-N as
detected in the {\em Hubble} Deep Field (HDF).  Shown is the full HDF
image, and circles show objects matched with X-ray sources in the 2 Ms
CDF-N. {\em (From Chandra press release; Credit: NASA/Penn State.)}}
\label{fopt}
\end{figure}

\chandra\ extragalactic surveys range from very deep and narrow to
shallow and very wide, allowing us to study the broadest possible
range in redshift and luminosity (Fig.~\ref{fsurv}).  The deepest
existing X-ray surveys are the Chandra Deep Fields (CDFs) North
\citep{alex03} and South \citep{luo08cdfs}, each with 2 Ms total
exposure.  Owing to \chandra's unparalleled spatial resolution, these
observations are not limited by confusion and probe to depths
more than 6 times fainter than is accessible with any other X-ray
observatory.  The CDFs have yielded extraordinary progress in
understanding faint X-ray populations and resolving the CXB.  However,
the X-ray luminosity density is dominated by more luminous X-ray
sources that are rare in the CDFs, so \chandra\ also has undertaken
several shallower surveys over wider areas to study these
objects. These surveys include, in order of increasing area, ELAIS-N
\citep{mann03elais}, the Extended CDF-S \citep{lehm05ecdfs}, AEGIS-X
\citep{nand05egs}, CLASXS and CLANS \citep{trou08optx} \citep[CLANS is comprised of data from the SWIRE/\chandra\ Survey;][]{wilk09sub}, C-COSMOS
\citep{elvi09sub}, XDEEP2 \citep{murr08head} and X\bootes\
\citep{murr05}.  The areas and corresponding flux limits for these
surveys are shown in Fig.~\ref{farea}.  Most surveys also have
extensive multiwavelength coverage with \hst, \spitzer, ground-based
optical imaging and spectroscopy, radio, and other observations
(Fig.~\ref{fopt}), allowing us to understand the detailed spectral
energy distributions (SEDs) and environments of the X-ray sources.
For a more detailed review of X-ray surveys with a focus on the deepest
fields, see \citet{bran05}.

In addition to dedicated \chandra\ observations in contiguous survey
fields, the \chandra\ Multiwavelength Project (ChaMP) uses archival
data for optical imaging and spectroscopic followup of serendipitous
sources from 392 (non-contiguous) ACIS pointings
\citep{gree04champ,kim07champ}.  The large samples nevertheless allow removal of
potentially biasing PI-targeted objects, and large statistical
subsamples immune to cosmic variance.  Similar follow-up of
serendipitous hard X-ray (2--10 keV) sources has been undertaken by
the Serendipitous Extragalactic X-ray Source Identification (SEXSI)
program \citep{harr03sexsi}.

\newpage
\noindent {\bf Resolving the cosmic X-ray background}

Since its discovery in rocket flights at the dawn of X-ray astronomy \citep{giac62},
the origin of the diffuse extragalactic X-ray background has been
one of the leading questions in high-energy astrophysics.  With the
exceptional sensitivity of the CDFs, it is now clear that almost all
of the extragalactic CXB at energies $<2$ keV arises from  X-ray point
sources.  The primary contributors are AGN, which dominate the
radiative energy density of the Universe at X-ray wavelengths.

The precise fraction of the CXB that is resolved in X-rays has been
the subject of numerous studies. \citet{more03} summed the X-ray
emission from sources detected in a variety of \chandra\ and other
surveys, and compared the total to an estimate of the total CXB,
concluding that $\approx$90\% of the CXB at $E<2$ keV was resolved.
\citet{wors05} performed a similar analysis as a function of energy,
and showed that the resolved fraction drops at energies $>5$ keV,
indicating a ``missing'' population of hard sources, which may include
AGN that are highly obscured by intervening gas (see below).  \citet{hick06a} took
a complementary approach, measuring the absolute flux of the {\em
unresolved} CXB in the CDFs, and showed that the resolved fraction of
the 1--2 keV CXB is $\approx80\%$.

 \citet{hick07b} further demonstrated that only $7\%\pm3\%$ of the 1--2 keV
CXB remained unresolved after excluding \hst\ sources in the GOODS
field, in broad agreement with a stacking analysis by
\citet{wors06}.  By studying the distribution of X-ray counts at the
\hst\ source positions, \citet{hick07c} showed that the \logn\ for
faint, unresolved X-ray galaxies in the CDFs is consistent with an
extension of the observed population of faint star-forming galaxies,
rather than AGN.  These results indicate a large population of faint
X-ray sources that may be accessible with deeper observations in the
CDFs.

\newsec
\noindent {\bf The cosmic evolution of black hole growth}

A key question in black hole evolution is: where and when did black
holes gain their mass?  Unlike galaxies (for which we can determine
ages for the stars), black holes have no ``memory'' of their formation
history.  Therefore, to determine the cosmic evolution of black hole
growth we must observe that growth directly, by measuring how the
space density of accreting black holes evolves with cosmic time.  A
number of authors have used the wealth of spectroscopic redshifts
available in X-ray surveys to derive the X-ray luminosity function for
AGN at a range of redshifts from the local Universe to $z>4$
\citep[e.g.,][]{ueda03, barg05, hasi05, silv08xlf}.  To cover the
largest possible region in the luminosity-redshift plane, these
studies combine data from narrow, deep and wide, shallow
\chandra\ surveys, as well as data from \xmmnewton\, \asca, \rosat, and other
missions.

\begin{figure}[!t]
\begin{center}
\epsscale{1.}
\plotone{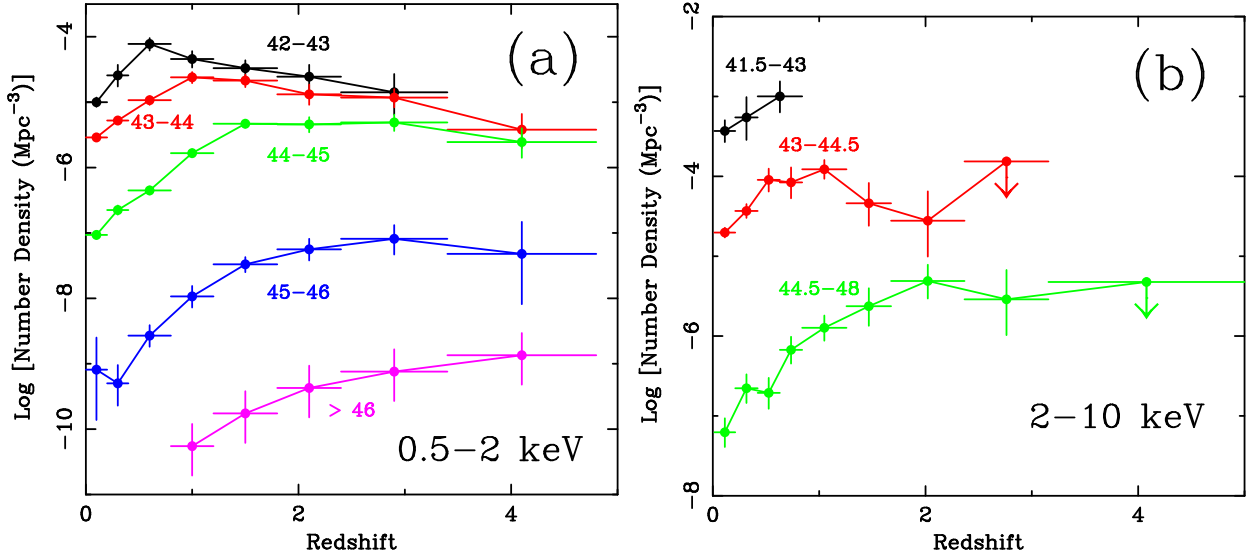}
\end{center}
\vspace{-6mm}
\caption{\footnotesize
The comoving space density of X-ray AGN as a function of redshift, shown in
the (b) 0.5--2 keV band \citep{hasi05} and (b) 2--10 keV band
\citep{ueda03}.  Lines show the evolution of AGN space density in
different bins of luminosity.  The data favor LDDE models, in which
the space density of low-luminosity AGN peaks at a lower redshift
compared to high-luminosity objects.  Figure compiled by
\citet{bran05}. \label{fxlf}}
\end{figure}

These studies have shown that AGN activity has a relatively complex
and interesting evolution with redshift.  \chandra\ results favor a
model of {\em luminosity-dependent density evolution} (LDDE), in which the
number density of AGN evolves differently for sources of varying
luminosities.  These results provide evidence for {\em downsizing}, in
which the density of the most luminous AGN peaks earlier in cosmic
time than for less luminous objects \citep[e.g.,][see
Fig.~\ref{fxlf}]{stef03agn, hasi05}, which can be shown to imply that
large black holes are formed earlier than their low-mass counterparts
\citep[e.g.,][]{merl08agnsynth, shan09agnbh}.  Qualitatively similar
downsizing has been observed for star formation in galaxies
\citep[e.g.,][]{cowi96sfev}, providing a circumstantial link between SMBH
and galaxy evolution.

\newsec
\noindent {\bf Understanding X-ray source populations}

The large numbers of AGN detected in extragalactic surveys allows for
robust statistical studies of AGN populations.  Particular effort has
been focused on measurements of X-ray spectra, which provide insights
into the nature of AGN accretion. \citet{tozz06} performed
spectral fits for hundreds of sources in the CDF-S, while
\citet{gree09champ} measured the spectra for $>1000$ SDSS quasars in
ChaMP, including 56 with $z>3$.  These studies show that in general,
the unabsorbed spectra of AGN have remarkably uniform power-law
continua.  However, \chandra\ studies  also have shown that X-ray
spectra and X-ray to UV SEDs of AGN get harder with decreasing
Eddington ratios\fnm\ \citep[e.g.,][]{stef06alphaox, kell08qsox}, similarly
to black hole X-ray binaries \citep{remi06araa}.

\fnt{The Eddington ratio is defined as the ratio of bolometric accretion luminosity to the Eddington limit, where  $L_{\rm Edd}=1.3\times10^{38} (M_{\rm BH}/M_{\sun})$ \ergs.}

X-ray surveys also can constrain the numbers of AGN that are absorbed
by intervening gas, which preferentially absorbs low-energy X-rays and
so hardens the observed spectrum.  The total spectrum of the CXB is
harder than the emission from a typical unabsorbed AGN, indicating a
significant contribution from absorbed sources.  In the local Universe,
75\% or more of optically selected Seyfert galaxies are obscured by
dust
\citep{maio95}, and \chandra\ surveys suggest a similar fraction are absorbed in X-rays.  Further, there is evidence that the obscured
fraction rises lower luminosities and may increase at higher redshifts
\citep[e.g.,][]{ueda03, stef03agn, hasi08agn}, which may constrain
models in which AGN provide radiation pressure feedback on surrounding
gas. In addition, X-ray studies have confirmed the identification of
obscured AGN detected in optical and IR observations
\citep[e.g.,][]{hick07abs, poll08obsc, donl08spitz, alex08compthick}.

Using the observed luminosity functions, spectral shapes, and
absorbing columns derived from \chandra\ and other surveys, it has
been possible to model the spectrum of the total cosmic X-ray
background \citep[e.g.,][]{trei05,gill07cxb}.  While these models
still have significant uncertainties (particularly in the number of
highly obscured AGN), their success implies that we may be converging
on a coherent picture for the cosmic evolution of black hole
growth. One caveat however, is that the low numbers of X-ray counts in
surveys make it difficult to distinguish absorption from intrinsically
hard spectra for faint sources. If the X-ray spectra of AGN become
harder at low Eddington ratio, this could produce a large number of
low-luminosity, X-ray hard AGN that would be classified as
``absorbed'' in current analyses \citep{hopk09lowlum}.  Future deep
surveys or detailed stacking of X-ray spectra may be able to break the
degeneracy between intrinsic spectral shape and absorbing column.

\begin{figure}[!p]
\epsscale{1.}
\plottwo{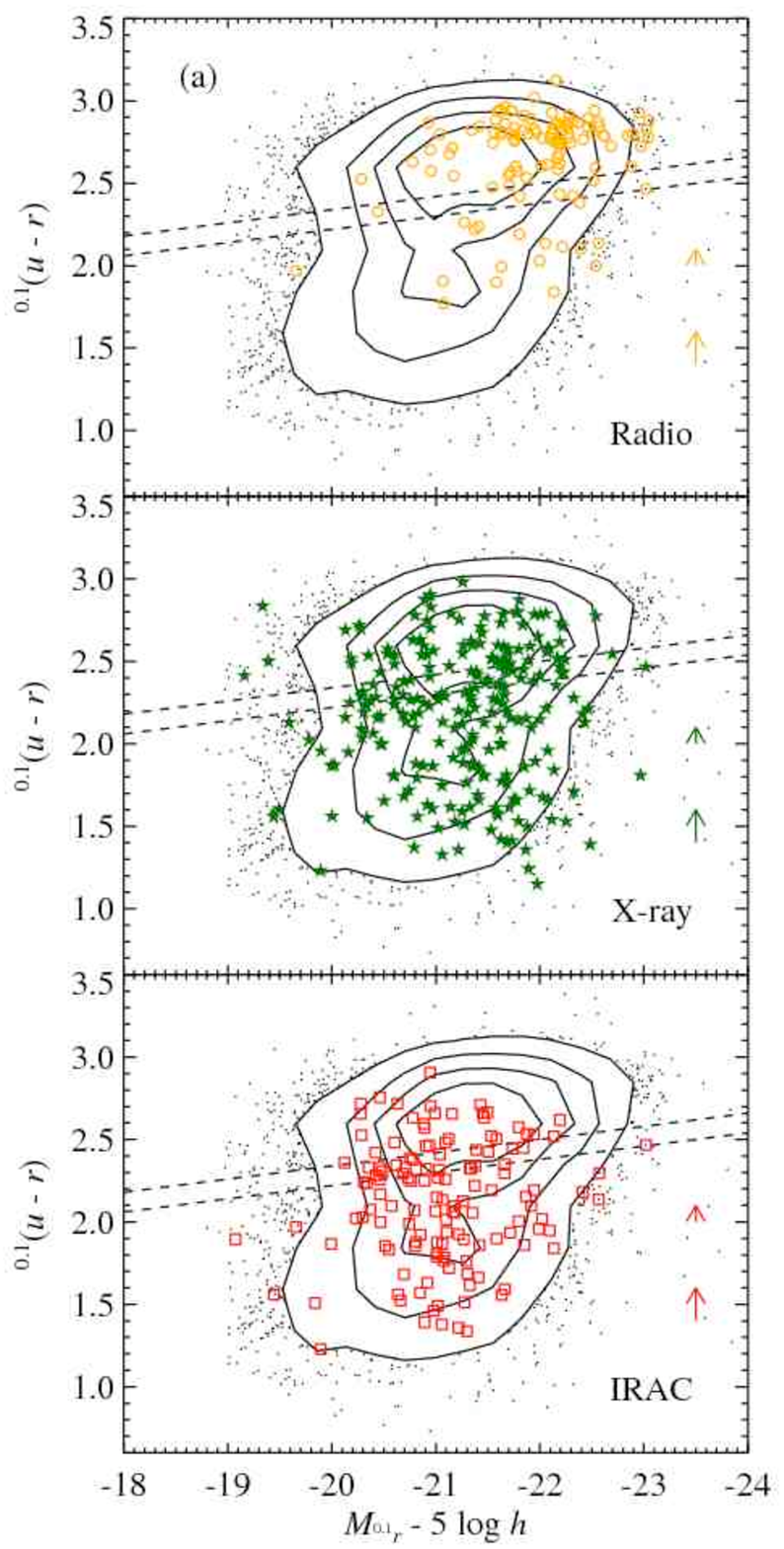}{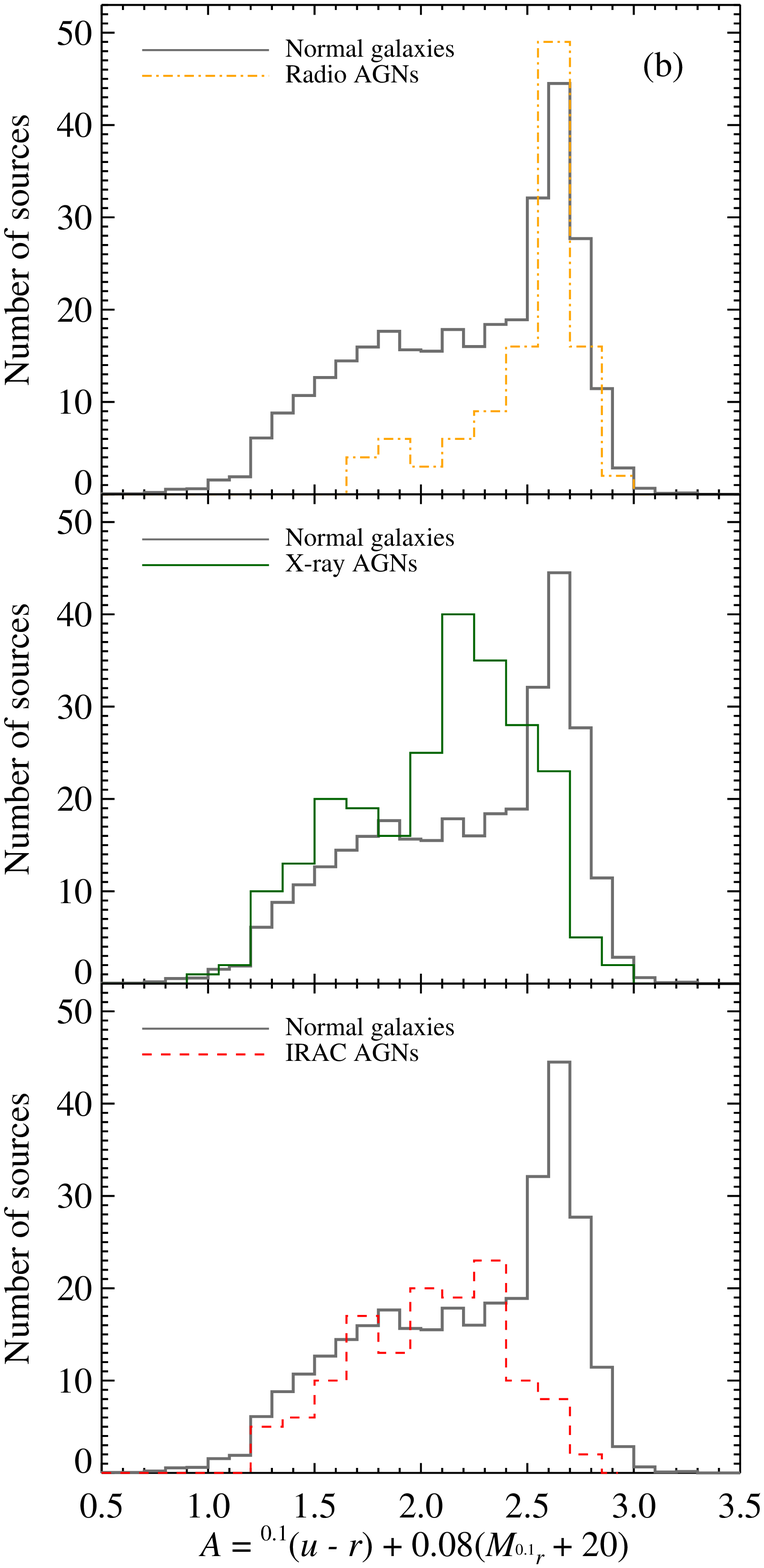}
\caption{\footnotesize({\em a}) Optical colors and absolute magnitudes
of AGNs at $0.25<z<0.8$ from the AGES redshift survey in \bootes\
\citep{hick09corr}.  Contours and black points show normal galaxies,
and lines separate red and blue galaxies.  The three panels show
radio, X-ray, and IR-selected AGNs, respectively.  ({\em b})
Distribution in rest-frame optical color for AGNs selected in the
three wavebands, compared to normal galaxies at $0.25<z<0.8$ (thick
gray line).  The radio AGN color distribution peaks along the red
sequence, while X-ray AGNs are found preferentially in the ``green
valley'' between the red sequence and the blue cloud.  The
distribution of IR AGNs is similar to that of X-ray AGNs, although
they are typically found in somewhat less luminous galaxies, and show
a less pronounced peak in the ``green valley''.
\label{fagncol}} \vskip0.5cm
\end{figure}

\newsec
\noindent {\bf Links between AGN and galaxy evolution}

Finally, \chandra\ surveys have allowed for detailed examination of
the host galaxies and environments and X-ray AGN, providing insight on
the role of AGN in the evolution of galaxies.  One powerful diagnostic
is the color-luminosity distribution for the galaxies that host AGN.
Galaxies are known to be divided into two types in color-magnitude
space: the red sequence of luminous, passively evolving galaxies, and
the blue cloud of less luminous, star-forming systems.  Interestingly,
a number of \chandra\ studies have found that at $z\lesssim 1$,
luminous X-ray AGN (unlike radio, optical, or infrared-selected AGN)
are preferentially found in the ``green valley'', with colors
intermediate between blue and red galaxies \citep[e.g.,][see
Fig.~\ref{fagncol}]{nand07host, geor08agn, silv08host, hick09corr},
indicating that X-ray AGN may be associated with the transition of
galaxies from the blue cloud to the red sequence, and may be
responsible for quenching the star formation in galaxies through
feedback \citep{bund08quench, geor08agn}.  Further, studies of the
environments and clustering of AGN find that \chandra\ X-ray AGN are
preferentially found in overdense regions characteristic of galaxy
groups \citep[e.g.,][]{yang06, geor07, silv09xcosmos_env, hick09corr,
coil09xclust}.  Models suggest that it is these environments where
star formation shuts off in massive galaxies; AGN may play a role in
the initial quenching of star formation, or in subsequent heating that
prevents gas from cooling and further forming stars
\citep[e.g.,][]{crot06, bowe06gal, hopk08frame1}.

\chandra\ surveys also have explored the presence of AGN associated with  
massive, vigorously star-forming galaxies in the distant Universe
$(z \sim 2)$.  X-ray studies of starburst galaxies selected in the CDFs
with observations in the submillimeter \citep[e.g.,][]{alex05} and
infrared \citep{dadd07comp, fior08agn} have provided evidence for a
large density of highly obscured AGN in galaxies co-eval with the
formation of the bulk of their stellar mass.  \citet{alex08bhmass}
found that the AGN in submm galaxies have black hole masses that are
roughly consistent with those expected from the local relation between black
hole mass and bulge mass, indicating that there may be continuous
feedback between star formation and accretion in these systems. While
the precise nature of the AGN population associated with luminous
starbursts is not yet clear, these studies point further towards
important links between the growth of galaxies and their central
SMBHs.

\newsec
\noindent {\bf The future}

Future X-ray surveys will provide more sensitive observations and
larger AGN samples to study the characteristics and evolution of SMBH
accretion in greater detail.  One future prospect with \chandra\ is
even deeper observations in the CDFs.  \chandra's high angular
resolution will allow it to observe significantly deeper (up to 8 Ms
or more) without reaching the confusion limit in the central regions
\citep{alex03}.  This would allow the detection of a new population of
extremely faint star-forming galaxies, as well as providing better
photon statistics for the sources that have already been resolved.
The upcoming NuSTAR\fnm\ mission (to launch 2011) will provide
\fnt{http://www.nustar.caltech.edu/. {\em Correction:} The original
version of this article stated that NuSTAR will perform an all-sky
survey; the mission will perform a series of smaller-area surveys.} an
unprecedented sensitive survey at hard X-ray (6--80 keV) energies,
while the eROSITA\fnm\ instrument on the {\em Spectrum X-Gamma}
observatory \fnt{http://www.mpe.mpg.de/projects.html\#erosita} (also
scheduled for 2011 launch) will survey the sky at 0.2--12 keV energies
and will provide enormous samples of X-ray AGN.  Among missions
proposed for the future, {\em Simbol-X}\fnm \fnt{
http://smsc.cnes.fr/SIMBOLX/} (target launch 2014) would provide high
angular resolution (better than 30\arcsec) and high sensitivity in the
$\sim$0.5--80 keV range, while EXIST\fnm
\fnt{http://hea-www.harvard.edu/EXIST/} would conduct a large-area
X-ray survey at very hard (5--600 keV) energies. The {\em Wide-Field
X-ray Telescope}\fnm \fnt{http://wfxt.pha.jhu.edu/} would provide an
analog to SDSS in the X-ray band, detecting $>10^7$ X-ray AGN over
$>10,000$ deg$^2$ in the energy range 0.1--4 keV (with the goal of
reaching sensitivity to 6 keV).  Further, the enormous sensitivity of
the {\em International X-ray Observatory}\fnm
\fnt{http://ixo.gsfc.nasa.gov/} would allow us to study in detail the
spectra of large numbers of faint AGN.  These future missions will be
essential to build on our understanding of the growth of black holes
and their place in the larger picture of galaxy and structure
formation in the Universe.

\noindent{{\em Acknowledgements:} Thanks to David Alexander, Francesca Civano, Bill Forman, Paul Green, Christine Jones, and Steve Murray for input, discussions, and comments on this article.  For the full Winter 2009 \chandra\ Newsletter and for previous editions, see http://cxc.harvard.edu/newsletters/.}

\newpage



\end{document}